\documentclass[aps,pra,twocolumn,superscriptaddress,showpacs]{revtex4-1}
\usepackage{graphicx}
\usepackage{amsmath,amssymb}

\newcommand{\vett}[1]{\mathbf{#1}}

\begin{document}

\title{Nonlinear optomechanical pressure}

\author{Claudio Conti} 
\affiliation{Department of Physics, University Sapienza, Piazzale Aldo Moro 5, 00185 Rome (IT)}
\date{\today}
\author{Robert Boyd} 
\affiliation{Department of Physics, University of Ottawa, 150 Louis Pasteur, Ottawa, Ontario, K1N 6N5 Canada}
\affiliation{Institute of Optics, University of Rochester, Rochester, New York, 14627, USA}
\affiliation{School of Physics and Astronomy, SUPA, University of Glasgow, Glasgow G12 8QQ, United Kingdom}

\begin{abstract}
A transparent material exhibits ultra-fast optical nonlinearity and is subject to optical pressure if irradiated
by a laser beam. However, the effect of nonlinearity on optical pressure is often overlooked, even if 
a nonlinear optical pressure may be potentially employed in many applications, as optical manipulation, 
biophysics, cavity optomechanics, quantum optics, optical tractors, and is relevant
in fundamental problems as the Abraham-Minkoswky dilemma, or the Casimir effect.
Here we show that an ultra-fast nonlinear polarization gives indeed a contribution to the optical pressure that also is negative in certain spectral ranges;
the theoretical analysis is confirmed by first-principles simulations. An order of magnitude estimate shows that the effect can be observable by measuring
the deflection of a membrane made by graphene.
\end{abstract}
\maketitle

The mechanical effect of light has been the subject of the investigations of many scientists for more than three centuries, as recently reviewed in \cite{marquardt2013}.
Foundational works have driven the emergence of fields of research, as, for example, optical tweezing and laser cooling \cite{Ashkin78}, quantum noise in interferometers \cite{Caves81}, and cavity optomechanics \cite{Dorsel83}.
Even if many aspects of opto-mechanical forces have been largely investigated, there are still several open 
problems, including, among others, the effect of the optical nonli\-nearity on the laser induced pressure.
As shown in recent papers  \cite{butsch2012,butsch:12a}, opto-mechanical deformations may produce huge optical nonlinearities, but if the opposite also holds true is at the moment unknown.

When considering a possible effect of an intensity dependent refractive index on the optical pressure, it is important to consider the issue of the form of the momentum of a photon in a dielectric. Indeed the related debate has characterized the li\-terature on 
optomechanical effects \cite{Brevik79, Milonni:10}.
The Abraham or the Minkoswky expressions of the photon momentum are to be chosen depending 
on the distinction between the canonical (i.e., the generator of translations) and the kinetic momentum \cite{Barnett:10}.
Even if in the absence of resonant light-matter interaction, it is accepted that the Abraham form is the correct one,
one may question what is the role of 
the always-present ultra-fast optical nonlinearity of electronic origin and, specifically, of an intensity-dependent refractive index.
The very same use of the Maxwell stress tensor, and the expression of the Abraham force may
also be questioned in the presence of nonlinearity.
This is an important issue in several fields,
including, among others, optical manipulation  \cite{dholakia:10,dileonardo:13,Lee:10,Chan:11,zemanek:13,saenz:13}, cavity optomechanics \cite{Kippenberg:08,marquardt2013}, 
biophysics \cite{russell2006,Russell2013,Garbos:11,Garbos:11a}, quantum optics \cite{pikovski12,Casimir:48, capasso:09},
optomechanics \cite{butsch2012,butsch:12a}.

The Balazs block (BB) furnishes a simple way for understanding the origin of the optical pressure\cite{balazs53}: a cubic piece of transparent matter obeying the Newton law in the absence of friction is irradiated by an electromagnetic (EM) wave
(see Fig.\ref{figurenum0}). 
As a photon travels through the block,
it is slowed down , and the block is displaced in the direction of light propagation.
Given the fact that the kinetic momentum of a photon is $\hbar \omega/(c n)$,
with $n$ the refractive index, $c$ the vacuum light velocity, $\omega$ the angular frequency,
and $\hbar$ the reduced Planck constant,
consider a photon that travels in vacuum, enters the block,
and after some propagation exits again in vacuum.
As the initial momentum for the photon is equal to the final one, 
the final BB velocity is zero.
However, as the momentum of the photon inside the BB ($n>1$) 
is smaller than in vacuum ($n=1$), 
during the passage of the photon the block moves to guarantee 
the momentum conservation, while the center of mass-energy travels at a constant velocity.
After the interaction, the BB 
is displaced by an amount proportional 
to its length $L$. Mechanical forces occur only at the entrance and at the exit of the photon.\cite{Testa2013}  
Albeit this analysis is oversimplified and hides a variety of fundamental problems,
and notwithstanding the fact that the BB displacement has never been observed in the experiments,
BB gives simple insights on optical pressure, due to forces arising from the interaction of the wave with the block interfaces.
The following arguments suggest a possible role of the optical nonlinearity.

Given the {\it linear} refractive index $n_0(\lambda)$ at the wavelength $\lambda$,
the photon velocity is $c/n_0$, i.e., the momentum divided by the ``mass'' $m$ obtained by the Einstein relation $mc^2=\hbar \omega$.
In the presence of an instantaneous nonlinear effect, the refractive index is $n=n_0+n_2 I$, 
$I(t)$ being the time-dependent intensity and $n_2(\lambda)>0$ the Kerr coefficient.
As the number of photons increases ($I$ increases), the velocity of the
photons in the block assumes the lower value $c/(n_0+n_2 I)$;
correspondingly, the BB momentum must increase, 
and a nonlinear contribution to the forces acting on the BB interfaces is be expected.
This may also be understood by noticing that 
the time-averaged optical force is determined by the optical transmission,
which is indeed affected by the nonlinearity.
For a single layer of transparent dielectric matter with length $L$,  
the power-dependent nonlinear phase shift alters the linear transmission due the Fabry-Perot effect and, correspondingly, the optical force.

In the following these arguments will be validated by a theoretical analysis, 
and by fully vectorial, four-dimensional, first-principles simulations of the nonlinear Maxwell equations.
We show below that the optical pressure is a quadratic function of the intensity because of nonlinearity, and that, in specific frequency intervals, the nonlinear contribution to the pressure is negative. This implies that for a reflection-less structure, a novel kind of all-optical tractor effect may be observable. Possible experimental tests could be performed with highly nonlinear
and mechanically resistant materials as, specifically, graphene\cite{Geim2007a}. 
We also discuss in the appendices, the modification of the Maxwell stress tensor
in the presence of an instantaneous nonlinear response.
\section{The effect of the nonlinear phase}
$\Sigma$ and $V$ being the surface area and the volume of the block, respectively, 
the time-dependent force due the EM wave is given by \cite{Stratton41}
\begin{equation}
\mathbf{F}=\frac{d \mathbf{G}_{mech}}{d t}=\int_\Sigma \bar{\mathbf S}\cdot\mathbf{\hat n} dA-\frac{1}{c^2}\frac{d}{d t}\int_V \mathbf{E}\times\mathbf{H} dV
\label{stratton}
\end{equation}
where $\bar{\mathbf{S}}\cdot\mathbf{\hat n}=\epsilon \mathbf{E}(\mathbf{E}\cdot \mathbf{\hat n})+\mu  \mathbf{H}(\mathbf{H}\cdot \mathbf{\hat n})- \small{\frac{1}{2}}(\epsilon E^2+\mu H^2) \mathbf{\hat n}$ is the projection of the Maxwell stress tensor 
$\bar{\mathbf{S}}$ on the unitary normal $\mathbf{\hat n}$ exiting from the surface $\Sigma$.
The last term in (\ref{stratton}) is the time-derivative of the electromagnetic (EM) momentum in the volume $V$,
whose density is $\mathbf{g}$, for which we adopted, following \cite{Stratton41}, the Abraham and von Laue expression $\mathbf{g}=\mathbf{g}_A=\small{\frac{1}{c^2}}\mathbf{E}\times\mathbf{H}$ (see also Appendix \ref{linearforce}). This corresponds 
to the kinetic momentum. \cite{Barnett:10}
In (\ref{stratton}) we neglect electrostrictive effects because of the fast time scale considered and of the known cancellation effects \cite{Milonni:10,Stratton41}.

In the continuous-wave (CW) case the average-force is given by $\overline {\bf F}=\small{\frac{1}{T}}\int_{-T/2}^{T/2} \mathbf{F}dt$, with the optical cycle $T=\lambda/c$, and $\overline {\bf F}$ the amount of momentum per unit time transferred to the block.
In the pulsed case,  the time-average per single pulse is defined as 
$\overline{\bf F}=\small{\frac{1}{T}}\int_{-\infty}^{\infty} \mathbf{F}dt$, which gives the total momentum transferred to the block per pulse during a normalization time $T$.

We consider a rectangular block with transverse dimensions $L_x$, $L_y$ and $L_z=L$ placed in vacuum with linear refractive index $n_0(\lambda)$ and Kerr coefficient $n_2(\lambda)$, as sketched in figure \ref{figurenum0}. 
In the theoretical analysis we assume an instantaneous optical Kerr effect, such that the overall refractive index is $n=n_0+n_2 I(t)$, with $I(t)$ the instantaneous intensity.
We analyze an $x$-polarized plane wave propagating in the $z$-direction.
The input (output) facet of the block is located at $z=0$ ($z=L$).

There is a very important issue to be considered when using the Maxwell stress tensor in the presence of nonlinear media. The general expression of $\bar{\mathbf{S}}$ in terms of electric and magnetic fields, and of the unitary dyadic $\bar{\mathbf{I}}$, reads as $\bar{\mathbf{S}}=\mathbf{D}\mathbf{E}+\mathbf{B}\mathbf{H}-\frac{\bar{\mathbf{I}}}{2}( \mathbf{D}\cdot\mathbf{E}+\mathbf{B}\cdot\mathbf{H})$, 
and is valid for a linear relation between $\vett{D}$ and $\vett{E}$ \cite{LandauEDCMbook,Milonni:10, Stratton41}. In the nonlinear case,
the overall force cannot be simply expressed as the flux of a tensor, but a volume integral that includes
the nonlinear polarization must be retained in the general case. 
As it happens in the isotropic case here considered, the properties of the nonlinear susceptibility tensor
may be such that the contribution of the nonlinear polarization may also be expressed as a surface integral and represented as an additive contribution to the stress tensor; this is detailed in the  Appendix \ref{nonlinearforce}.

In the following, we calculate the lowest order contribution to the optomechanical force in $n_2$.
Letting $\mathbf{\hat z}$ be the unit vector co-directional to the $z$-direction, we have from (\ref{stratton}) for the longitudinal z-components $G_{mech,z}=\mathbf{G}_{mech}\cdot \hat {\bf z}$, and $F_z=\mathbf{F}\cdot \hat{\mathbf z}$. Letting $A=L_x L_y$ the transverse block area, the optical pressure $p$ is here defined as the time-averaged force per unit of area, as obtained by the flux of the Maxwell stress tensor. Note that this is not a true pressure (i.e., independent of the direction), but is the force per unit of surface in the direction of the wave propagation. 
Neglecting transverse directions,
\begin{equation}
p=\frac{\overline{F_z(t)}}{A} = \overline{\frac{I(0,t)}{c}} - 
\overline{ \frac{I(L,t)}{c}}\text{,}
\label{pressure}
\end{equation}
where $I(z,t)$ is the {\it instantaneous} optical intensity.
We find that the optical pressure can be expressed as a 
\begin{equation}
p=p_1 I_0+\small{\frac{1}{2}} p_2 I_0^2
\label{pressureKerr}
\end{equation}
being $I_0$ the peak intensity and $p_2$ a coefficient, denoted hereafter as the ``nonlinear pressure coefficient'', which vanishes in the absence of the optical Kerr effect. In the following we give below the expressions for $p_1$ and $p_2$ for CW and pulsed optical excitation.
\section{Continuous wave excitation}
Given the input intensity in the CW case $I(0,t)=I_0 \sin(\omega t)^2$, with $\omega=2\pi c/\lambda$ the optical angular frequency, 
the EM propagation through the block induces a linear $\phi_L=2\pi n_0(\lambda) L/\lambda$ and a nonlinear phase-shift $\phi_{NL}(t)=2\pi n_2(\lambda) I(0,t) L/\lambda$. The intensity at the output is hence
\begin{equation}
I(L,t)=\mathcal{T}(\lambda) I_0 \sin\left[\omega t-\phi_L-\phi_{NL}(t)\right]^2\text{.}
\label{iLcw}
\end{equation}
In (\ref{iLcw}), $\mathcal{T}(\lambda)$ is the linear Fabry-Perot transmission from the block, also including linear absorption losses.
A direct calculation by Eq.(\ref{pressure}) gives
\begin{equation}
p_1=\frac{1-\mathcal{T}(\lambda)}{2 c}\text{,} 
\label{p1}
\end{equation}
which is a known results, as outlined, e.g., in \cite{Milonni:10}. 
Eq.(\ref{p1}) shows that the leading part of the optical pressure is due to finite transmission, and 
vanishes for an index-matched system with no absorption, i.e., for $\mathcal{T}=1$. 

For the nonlinear part we have
\begin{equation}
p_2\equiv p^{CW}_2(\lambda)= \frac{\pi \mathcal{T}(\lambda)  n_2(\lambda) L }{c \lambda}  \sin\left[\frac{4 \pi n_0(\lambda) L}{\lambda}\right]\text{.}
\label{p2cw}
\end{equation}
Eq.(\ref{p2cw}) implies that $p_2$ can be either positive or negative depending
on the wavelength and on the size of the block.

It is remarkable that a perfectly matched device, such that $\mathcal{T}=1$ and $p_1=0$, sustains in specific spectral ranges a {\it negative} optical pressure, i.e., $p=-\small{\frac{1}{2}}|p_2|I_0^2<0$, resulting in a {\it tractor effect}.
For $\mathcal{T}<1$, Eq.(\ref{p2cw}) predicts the existence of an intensity $I_{tractor}$, such that $p$ changes from positive to negative values, i.e., for  $I_0=I_{tractor}=2 p_1/|p_2|$, a transition that occurs if $p_2<0$ for a given block length $L$ and wavelength $\lambda$. 
$I_{tractor}$ is directly proportional to the reflection 
coefficient, and hence the transition is potentially observable for nearly index-matched blocks. We also mention the possibility of using the Brewster angle to maximize transmission and attain negative optical pressure.
It is very important, however, to underline that the exact condition $\mathcal{T}=1$ is not physically realizable, 
even in the absence of absorption, because of finite transverse size effects and because a purely monochromatic wave is an idealization.

In figure \ref{figuretheory} we summarize the leading features of this theoretical analysis with reference
to realistic parameters: Fig.\ref{figuretheory}A shows the considered linear and nonlinear dispersion of 
$n_0$ and $n_2$ (the details of the adopted model are given in the Appendix \ref{codedetail}); Fig.\ref{figuretheory}B shows the pressure per unit peak intensity $p/I_0=p_1+p_2 I_0/2$ for $L=2~\mu$m;
Fig.\ref{figuretheory}c shows the nonlinear pressure coefficient $p_2$ in terms of lambda $\lambda$, after Eq.(\ref{p2cw}).
In the Appendix \ref{delayed} we discuss the effect of a non-instantaneous nonlinear response.
\section{Pulsed wave excitation}
For a pulsed excitation, we take the input intensity Gaussianly modulated in time:
\begin{equation}
I(0,t)=\sqrt{2}I_0 e^{-\left(\frac{t-t_s}{T_0}\right)^2}\sin(\omega t)^2\text{,}
\label{i0pulse}
\end{equation}
and the normalization time is chosen as $T=\sqrt{2\pi} T_0$; in this way, Eq.(\ref{i0pulse}) is such that the time-averaged intensity $\overline I(0,t)$ is equal to the CW case considered above.
The pulse given by (\ref{i0pulse}) disperses during propagation; we only consider the effect of the group delay, and neglect
second and higher order dispersion by assuming $L$ small with respect to the dispersion length.
As shown below, for group delays (long propagation) much greater then the pulse width $T_0$ the nonlinear pressure becomes negligible.
The inverse group velocity is $1/v_g=n_0/c+\omega n_0'/c$, 
the transmitted pulse exhibits phase $\phi_L=2\pi n_0(\omega) L/\lambda$ and
 group delay $t_g=L/v_g$, and a nonlinear phase shift $\phi_{NL}$:
\begin{equation}
I(L,t)=\mathcal{T}(\lambda) \sqrt{2} I_0e^{-\frac{(t-t_s-t_d)^2}{T_0^2}}\sin[\omega t-\phi_L-\phi_{NL}(t)]^2\text{.}
\label{ilpulse}
\end{equation}
As above, in (\ref{ilpulse}), $\phi_{NL}(t)=2\pi n_2 I(0,t) L/\lambda$. By using these definitions, 
we find, after a direct calculations, the expressions for $p_1$ and $p_2$. The general 
result is cumbersome and is given in the Appendix \ref{pulsedall}. For pulse duration $T_0$ greater than a few optical cycles,
(i) the linear coefficient $p_1$ is identical to Eq.(\ref{p1}), and
(ii) the nonlinear contribution to the optical pressure is given by
\begin{equation}
p_2(\lambda,T_0,t_d)=p_2^{CW}(\lambda)\,\exp{\left(-\frac{{L}^2}{2 v_g^2 T_0^2}\right)}
\label{p2pulse}
\end{equation}
with $p_2^{CW}$ given by Eq.(\ref{p2cw}). 
The nonlinear pressure is hence a function of the pulse duration and of the group delay, as given by Eq.(\ref{p2pulse}).
For $T_0\rightarrow\infty$ (very long pulses with respect to the group delay) the $CW$ expression (\ref{p2cw}) is re-obtained.
On the other hand, the nonlinear pressure vanishes for propagation times $L/v_g$ much longer than the pulse duration.
Figure \ref{figuretheory}D shows the ratio $p_2/p_2^{CW}=\exp(-L^2/2 T_0^2 v_g^2)$ after Eq.(\ref{p2pulse}).

\section{Nonlinear Maxwell equations}
We validate the theoretical analysis by a first-principles numerical approach,
the Finite Difference Time Domain (FDTD) algorithm \cite{TafloveBook}.
We solve the full Maxwell equations including dispersion in the linear and nonlinear material response.
The model considered in the simulations is much more general than the theoretical
analysis above. Specifically: (i) transverse effects, as finite beam size and boundary effects at the
block lateral surfaces, are included; (ii) nonlinearity is not exactly instantaneous, but follows textbook models \cite{BoydBook,BloembergenBook} for the ultra-fast electronic third-order
susceptibility; (iii) the calculated transmission function includes dispersion, nonlinear effects,
 multiple reflections for the pulsed case, and we also introduce a not-negligible amount of linear absorption (see Appendix \ref{codedetail}). 
In these respects, the numerical simulations allow us to strictly test the validity of the theoretical analysis.
Previously, other authors calculated by FDTD techniques the optical pressure on dielectric media \cite{mahajan:03,Zhang:04,Gauthier:05,Moloney:05,Jiang:06,Sung:08} (see also \cite{johnson:13} for other approaches); to the best of our knowledge nonlinearity has not been considered before.

We simulate in three spatial dimensions (3D) and time the nonlinear Maxwell equations
\begin{equation}
\begin{array}{l}
\nabla\times\mathbf{E}=-\mu_0 \partial_t\mathbf{H}\\
\nabla\times\mathbf{H}=\partial_t \mathbf{D}_{L}+\partial_t \mathbf{P}_{NL}\text{,}
\end{array}
\end{equation}
with $\mathbf{D}_{L}$ the linear displacement vector, which is given by  $\mathbf{D}_{L}(\omega)=\epsilon_0 \epsilon_r(\omega) \mathbf{E}(\omega)$
in the frequency domain, with linear refractive index $n_0(\omega)=\sqrt{\epsilon_r(\omega)}$. Nonlinearity is given by a nonlinear oscillator with a Kerr coefficient $n_2(\omega)$
of the order of $10^{-22}~$m$^2$/W 
Linear and nonlinear material dispersion are given by a single pole model and shown in figure \ref{figuretheory}A; 
further details are in the Appendix \ref{codedetail}. 

\begin{figure}
\includegraphics[width=0.45\textwidth]{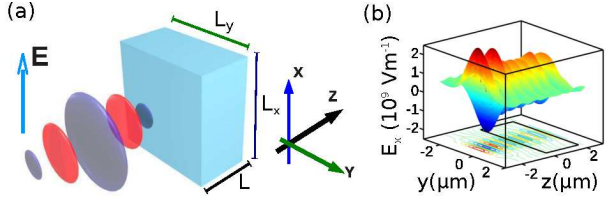}
\caption{(Color online) (a) Sketch of the simulated structure, the input field and the block are indicated; (b)
snapshot from one typical FDTD simulations in CW of the field $E_x$ during propagation in the ($y$,$z$) section at
$t=65$~fs and $P_0=95~$kW. The BB  position is indicated by the black thick-line in the horizontal plane.
\label{figurenum0}}
\end{figure}

The geometry of the simulations is sketched in figure \ref{figurenum0} and we analyze CW and pulsed excitation.
We consider a block with sizes $L_x=L_y=4$~$\mu$m, and $L_z=L=2~\mu$m.
The input beam is a linearly polarized $TEM_{00}$ Gaussian beam with waist $w_0=1~\mu$m located at the entrance facet of the block and wavelength $\lambda=800$~nm. 
In the simulations we change the input peak power of the beam $P_0$. Figure \ref{figurenum0}b shows a snapshot of the $E_x$ component of the field in the $(y,z)$ plane in the simulated geometry. 
\begin{figure}
\includegraphics[width=0.45\textwidth]{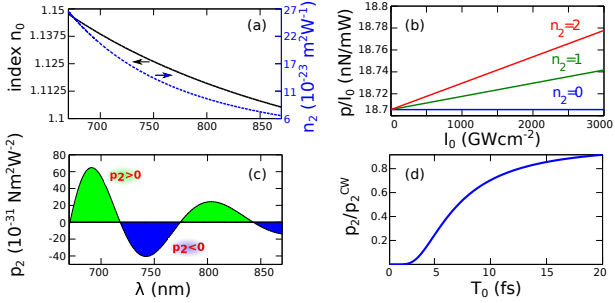}
\caption{(Color online) (a) Dispersion of the linear refractive index $n_0(\lambda)$ (left scale) and the nonlinear Kerr coefficient $n_2(\lambda)$ (right scale) used for the
theoretical analysis and in the numerical simulations (the model is detailed in the Appendix \ref{codedetail}) ;
(b) pressure per unit of intensity (force per Watt) for various $n_2$ in units of $10^{-22}\,m^2/W$;
(c) nonlinear pressure coefficient versus wavelength, with $n_2$ as in panel (a);
(d) nonlinear pressure coefficient versus pulse duration ($L=2~\mu$m, $\lambda=800$~nm).
\label{figuretheory}}
\end{figure}
\begin{figure}
\includegraphics[width=0.45\textwidth]{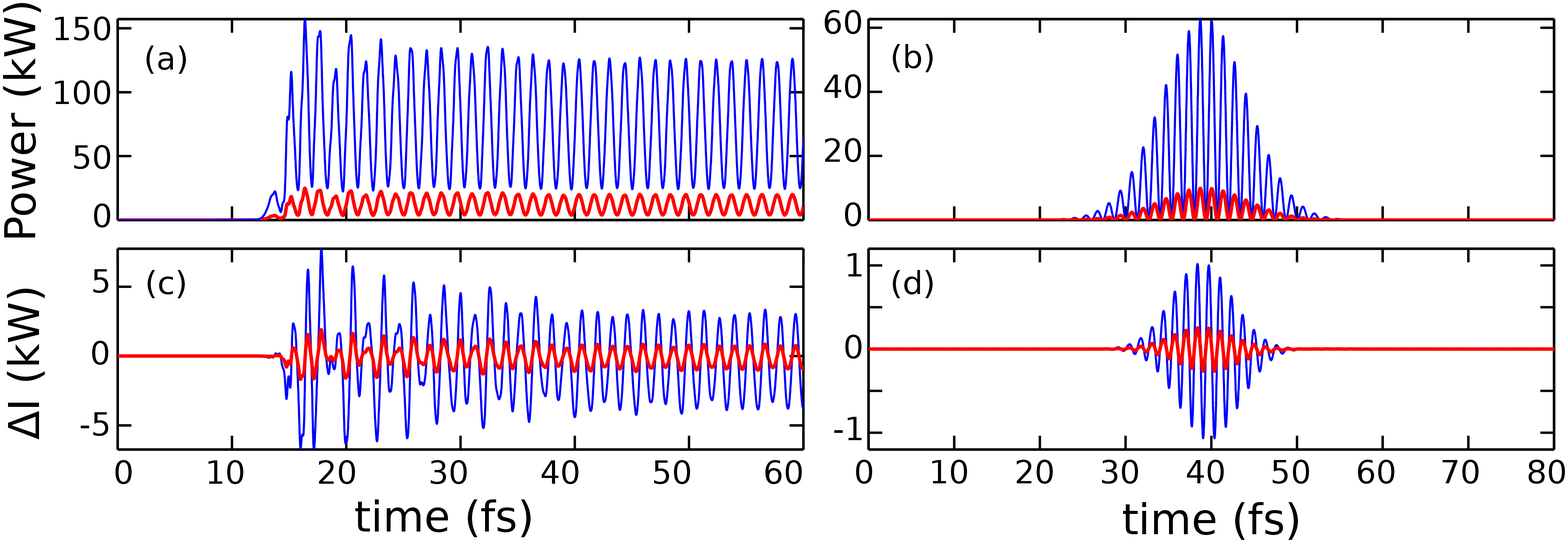}
\caption{(Color online) (a) Transmitted instantaneous power in the CW case for input peak power $P_0=15$~kW (thick red line) and  $P_0=95$~kW (blue line);
(b) as in (a) with pulsed excitation with pulse duration $T_0=10$~fs;
(c) $\Delta I$ after Eq.(\ref{DeltaI}) for $P_0=95kW$ and $n_2\cong 10^{-22}$~m$^2$/W (thick red-line) and $n_2\cong 2\times 10^{-22}$~m$^2$/W (blue thin line);
(d) as in (c) for pulsed excitation with $T_0=10$~fs;
\label{figurenum2}}
\end{figure}

In figure \ref{figurenum2}A,B we show the output flux of the Poynting vector for the CW and pulsed cases, and for two different peak powers. 
In the CW case there is an initial transient needed by the input wave to travel through the BB.

The nonlinear phase shift alters the BB transmission; this is simply revealed in the simulations by calculating the difference $\Delta I(t)$ between the transmitted intensity $I_L(t)=I(L,t)$ for $n_2>0$ and that obtained in the linear regime by letting $n_2=0$, denoted as $I_{L,n_2=0}(t)$. At the lowest order in $n_2$, we have
\begin{equation}
\Delta I=I_L-I_{L,n_2=0}=I_0 \phi_{NL}(t) \sin(2\phi_L-2\omega t)\text{.}
\label{DeltaI}
\end{equation}
$\Delta I$ is a signal with carrier $2\omega$ and amplitude modulation given by $\phi_{NL}(t)$.
In figure \ref{figurenum2}C,D we show $\Delta I(t)$ for two different values of $n_2$ and the same input power; the amplitude modulation of $\Delta I$ grows
 with the amount of nonlinearity. The origin of this modulation in the transmitted intensity is the nonlinear phase-shift that alters the Fabry-Perot effect.

The force components are calculated as the 3D flux of the Maxwell stress tensor
over the entire surface of the block, thus including transverse effects due to polarization, 
the finite size of the beam and of the block, and the material dispersion in the linear and nonlinear response. The transverse components $F_{x,y}(t)$ of the force (not reported) are found to be
orders of magnitude smaller than the longitudinal force $F_{z}(t)$.
\begin{figure}
\includegraphics[width=0.45\textwidth]{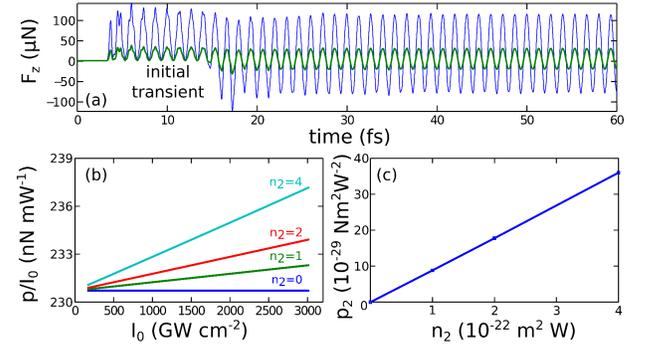}
\caption{(Color online) (a) Calculated force $F_z(t)$ for two input peak powers $P_0=25$~kW (thick green line) and $P_0=95$~kW (blue thin line).
Note the initial transient regime for $t<15$~fs needed for the wave to travel within the block ($n_2\cong 10^{-22}$~m$^2$/W); 
(b) the pressure per unit intensity calculated for various values of $n_2$ indicated in the panel and given in units of $10^{-22}~$m$^2$/W (c);
the pressure is calculated as the time-average in the stationary regime $t>40$~fs; 
(c) nonlinear pressure coefficient $p_2$ as determined from data in panel (b) for various $n_2$;
\label{figurenum4}}
\end{figure}
\begin{figure}
\includegraphics[width=0.45\textwidth]{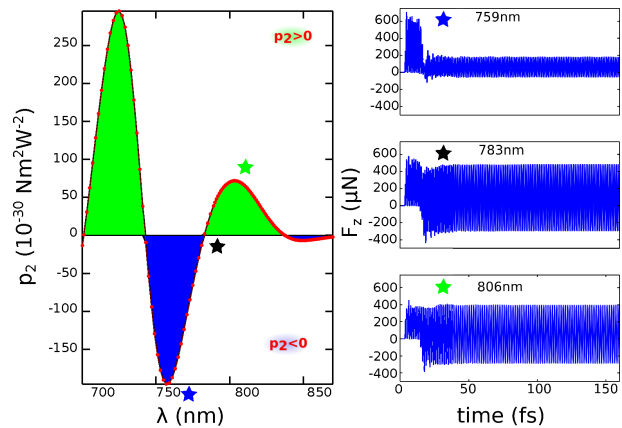}
\caption{(Color online) Left panel, $p_2$ coefficient versus input wavelength for a CW excitation ($n_2\cong 10^{-22}$~m$^2$/W); note the region of negative $p_2$. Panels on the right show the trend of the calculated force $F_z(t)$ for specific wavelengths, as indicated.
\label{figurenum5}}
\end{figure}
Figure \ref{figurenum4}A shows the time dynamics of the force $F_z(t)$ for the CW excitation;
the input signal is a sinusoidal function, and after an initial transient needed for the wave
to fill all the block ($t<15$~fs in Fig.\ref{figurenum4}A), a stationary regime is reached.
In figure \ref{figurenum4}A we show the calculated force for two values of the input peak power.
Figure \ref{figurenum4}B  shows the resulting pressure as defined in Eq.(\ref{pressure}) divided 
by the optical peak intensity for various values of the
nonlinear coefficient $n_2$. A nonlinear contribution to the pressure is present.
The calculated $p_2$ coefficient versus $n_2$ is shown in figure \ref{figurenum4}C and follows Eq.(\ref{p2cw}).

To determine $p_2$, as defined by equation (\ref{pressureKerr}) we perform several simulations by varying intensity $I_0$,
and calculate the resulting time dependent force $F_z(t)$ from the flux of the Maxwell stress tensor over the whole surface of the block.
$F_z(t)$ is divided by the by the area $A=L_x L_y$, and averaged with respect to time, this determines the function $p(I_0)$; 
$p_2$ is numerically calculated as the second derivative $p_2=d^2 p/d I_0^2$. 
In the continuous case $p$ is calculated by averaging the temporal signal $F_z(t)$ obtained by the FDTD simulation over an optical cycle,
to avoid the initial transient we consider the time profile for $t>40$~fs.
In the pulsed regime below $F_z(t)$ is integrated over the whole temporal axis and divided by $\sqrt{2\pi} T_0$ as described in the text.
This procedure is repeated for all the considered wavelengths, pulse durations, and nonlinear coefficients.


We also numerically investigated the dependence of the nonlinear pressure on the input wavelength 
as shown in figure \ref{figurenum5}; it follows the trend predicted by Eq.(\ref{p2cw}),
 in Fig.\ref{figuretheory}c.
We remark the existence of specific frequencies where $p_2$ vanishes, and spectral regions where the nonlinear pressure coefficient is negative. Discrepancies in the spectral distribution of $p_2$ in the simulations and in the theory  are ascribed to the fact that in the simulated nonlinear Maxwell equations the nonlinearity is not exactly instantaneous, and to the linear losses included in the simulated model (see Appendix \ref{codedetail}).

Figure \ref{figurenum6} shows the instantaneous force for pulsed excitation
($T_0=10$~fs, $P_0=70$~kW). The input impulse due to the first interface of the block,
 and the opposite one at the block exit are indicated in figure \ref{figurenum6}A.
Figure \ref{figurenum6}B shows the calculated pressure $p$, including the linear and nonlinear parts, for various
input pulse duration $T_0$ in panel \ref{figurenum6}C. The latter shows the calculated trend of the
nonlinear pressure coefficient versus $T_0$, which follows Eq.(\ref{p2pulse}).

\begin{figure}
\includegraphics[width=0.45\textwidth]{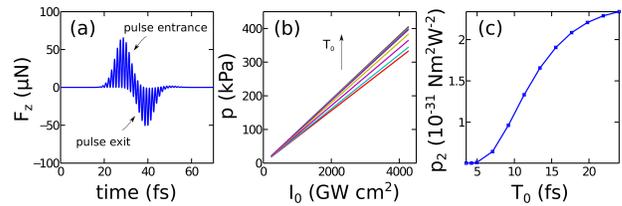}
\caption{(Color online) (a) Calculated force $F_z(t)$ versus time in the presence of a pulsed excitation ($T_0=10$~fs, $P_0=70$~kW);
note that the force is at a maximum in correspondence of the entrance and of the exit of the pulse from the BB;
(b) pressure $p$ versus input intensity $I_0$ for the various pulse duration corresponding to the dots in panel (c); 
(c) nonlinear pressure $p_2$ versus pulse duration $T_0$ at fixed power $P_0=70$~kW and $n_2\cong 10^{-22}$~m$^2$/W.
\label{figurenum6}}
\end{figure}

\section{A graphene optical sail}
The nonlinear contribution to the optical force is expected to play a role is several different frameworks;
but as a first analysis a material that could be used to experimentally measure a nonlinear opto-mechanical force
should exhibit a large nonlinear optical response and be available in thin layers able to sustain relevant mechanical 
stress. In these respects, graphene looks to be a very interesting candidate \cite{Geim2007a}. 
For example, one could consider the mechanical deformation
of a graphene membrane anchored at the boundaries and irradiated by an intense laser beam.
Graphene is one the strongest known materials and is hence very well suited to sustain large optomechanical stresses. 

A possible experimental geometry could be that used in \cite{Poot2008} to measure the elastic properties of 
graphene thin layers: circular membranes are suspended in the holes of a substrate and deformed by 
atomic force microscopy (AFM) nano-sized cantilevers. Here, instead of AFM nanoindentation, we consider the case in which the displacement is induced by a focused laser beam.  For a beam waist $w_0=10~\mu$m, wavelength $\lambda=532$~nm, and optical power $P=20$~W, we consider a circular membrane with radius equal to $w_0$, so that the optical pressure is uniform over the surface. 
We remark that this configuration is different from the case of the AFM probe, as the force is not localized in the center of the membrane, but involves its entire area.
Correspondingly, the maximum vertical displacement $W$ of the graphene layer can be calculated by \cite{AmenzadeBook}
\begin{equation}
W=\frac{p w_0^4}{64 D}\text{,}
\label{displacement}
\end{equation}
being $D$ the bending rigidity. For graphene layer width $L=10~$nm, $D=10^{-13}$~N~m \cite{Poot2008}.
As the deflection $W$ at the mechanical equilibrium grows with the optical pressure, a nonlinear optical contribution result in a variation of the spatial deformation, a kind of {\it optical sail}. 

Graphene has a linear complex refractive index $2.6-i 1.3$,\cite{Geim2007}, and linear absorption can be neglected for the considered 
small values of $L$; Fabry-Perot thin-film reflectivity for $n_0=2.6$ is of the order of $10\%$ ($\mathcal{T}=90\%$).
Without including nonlinearity, the pressure is $p=p_L=p_1 I_0$ after Eq.(\ref{pressureKerr}), 
and the mechanical force $F_L=\pi w_0^2 p_L\cong 3$~nN, much lower than
the measured maximum sustainable breaking values \cite{Poot2008}. $F_L$ induces a displacement $W\cong W_L=17$~nm after Eq.(\ref{displacement}).

When including nonlinear optical effects, we have from Eq.(\ref{pressureKerr})
$p=p_{NL}=p_1 I_0 +(1/2)p_2 I_0^2$, and relative variation
\begin{equation}
\frac{\Delta p}{p_L}\equiv\frac{p_{NL}-p_L}{p_L}=\frac{p_2 I_0}{2 p_1}=\frac{\mathcal{T}}{1-\mathcal{T}}\frac{4 \pi^2 L^2 n_0 n_2 I_0}{\lambda^2}\text{,}
\label{dff}
\end{equation}
as found by using Eqs.(\ref{p1}, \ref{p2cw}), and under the hypothesis of a very thin layer, such that the sine function in (\ref{p2cw}) can be approximate by its argument. 
In the considered case 
\begin{equation}
\frac{\Delta p}{p_L}\cong 0.3 n_2 I_0 \text{.}
\label{dff1}
\end{equation}
Graphene has giant nonlinear optical response $n_2=10^{-7}~$cm$^2$~W$^{-1}$, and the 
considered intensity $I_0=P/(\pi w_0^2)\cong 0.01$~GW~cm$^{-2}$
induces a nonlinear refractive index correction $n_2 I_0\cong 0.6$; 
this fluence level is such that nonlinear absorption is negligible.\cite{Zhang:12}
Eq.(\ref{dff1}) implies that a few layers of graphene exhibit a relative optical pressure variation of the order of $20\%$ at a moderate intensity level due to nonlinear effects. 
The corresponding force is $F_{NL}=\pi w_0^2 p_{NL}=4~$nN, 
and the resulting deflection, following Eq.(\ref{displacement}), is $W\cong W_{NL}=20~$nm.
The small variation of $W$ due to the nonlinear contribution of the optical pressure is of the order of ten graphene layers, and looks within the range of measurable displacements by the techniques so far employed. This suggests that the effect of the nonlinear optical pressure may be observable in a simple experiment by graphene.

\section{Conclusions}
In conclusion we have theoretically shown that nonlinearity affects the opto-mechanical force.
The results have several possible implications as, for example, investigating the 
kind of mechanical forces arising from nonlinear waves as spatial solitons, optical bullets or rogue waves.
We considered the simplest ultra-fast Kerr effects, but issues such as spatial non-locality,
delayed temporal responses, wave-mixing among polarizations or spectral frequencies may be analyzed in the future.
The whole set of spatio-temporal effects that may also arise when considering spatial shapes more complicated than 
a simple cubic box may also affect the opto-mechanical forces, e.g., focusing actions inside spheres may enhance 
the nonlinear pressure. The fact that the nonlinear contribution to the force may be negative open several possible
roads of investigations in terms of the optimization and the enhancing of ultra-fast broad band tractor effects, by using, for example, 
pulse-duration, spatial and polarization shaping, and wavelength mixing. 
Other kinds of nonlinearity could be considered, as quadratic parametric interactions and self-induced transparency, 
and the possibility of having multiple effects also in spatially non-homogeneous systems let us envisage that the nonlinear force may
have a substantial role in practical applications.
Last but not least, frequency mixing phenomena and super-continuum generation in nonlinear systems do open a variety of fundamental problems in terms
of the momentum exchange mediated by photons in moving media, which are also important in the fully quantum regime,
where different states and squeezing of light in the presence of nonlinearity may largely affect opto-mechanical motion in many at the moment still unknown possibilities.
A simple order of magnitude analysis shows that graphene could be the perfect material to investigate the opto-mechanical pressure with nonlinear origin, 
as this material displays a huge optical nonlinear response and has the required mechanical and thermal properties to sustain high power laser beams,
also in the continuous wave regime. This opens the way to a variety of further possible applications.

\begin{acknowledgments}
We gratefully acknowledge fruitful discussions with Philip Russell, support from the Humboldt foundation,
and the hospitality of the Max Planck Institute for the Science of Light.
RWB gratefully acknowledges support from the Canada Excellence Chair Program.
CC acknowledges support from the Sapienza research project 2012 SUPERCONTINUUM,
and from the COMPLEXLIGHT ERC project (grant number 201766).
The numerical work reported in this manuscript has been developed within the Italian Supercomputing Resource Allocation (ISCRA) at the CINECA and the parallel simulations were performed on the IBM Blue Gene Q system FERMI. 
\end{acknowledgments}

\appendix
\section{\label{linearforce}Optical force in the linear case}
For the sake of completeness we start recalling the basic theory of the Maxwell stress tensor,
following the notation of \cite{Milonni:10}, and with reference to the linear case.
In our model, a material medium is treated as a distribution of dipoles with polarization $\vett P$. 
The polarization in the linear case obeys the equation\cite{TafloveBook}
\begin{equation}
\frac{d^2 \vett{P}}{d t^2}+2\gamma \frac{d \vett{P}}{d t}+\omega_0^2 \vett{P} =\omega_0^2 \epsilon_0 
\left[\epsilon_s(\vett{r})-1 \right] \vett{E} \text{.}
\end{equation}

In the presence of an external electric field, the dipoles are subject to the Lorentz force acting on their charges
and their displacement current $\vett J=\partial_t \vett P$. 
The charge $\rho$ is given by the continuity equation $\partial_t \rho=\nabla\cdot \vett J$, and also 
$\rho=\nabla\cdot \vett D=\nabla\cdot (\epsilon_0 \vett E+\vett P)$.

The force volume density is $\vett f_L=\rho \vett E+\vett J\times \vett B$, and integrated on a volume $V$ that strictly contains all the charges, with surface $\Sigma$, gives the force acting on the medium
(neglecting surface effects, \cite{Stratton41, Gordon73})
\begin{widetext}
\begin{equation}
\vett F_L=\int_V \vett f_L dV=\int \rho \vett E+\vett J\times \vett B dV=\int \nabla\cdot \vett (\epsilon_0 \vett E+\vett P) \vett E+\vett \partial_t \vett P\times \vett B dV\text{.}
\label{FL1}
\end{equation}
\end{widetext}

In the frequency domain we have $\vett{\tilde{P}} (\omega)=\epsilon_0 \chi^{(1)}(\omega) \vett{\tilde E}$, with $\chi^{(1)}$ given in Appendix E,
and being the relative dielectric permittivity $\epsilon_r=1+\chi^{(1)}=n^2(\vett r)$, with $n(\vett r)$ the refractive index.
From Eq.(\ref{FL1}) one has
\begin{equation}
\vett F_L=\vett F^A+ \int_V \epsilon_0 (n^2-1) \frac{1}{2}\nabla (E^2) dV=\vett F^A+\vett F_M\text{,}
\label{FL2}
\end{equation}
which is the well known expression for the optical pressure on a linear medium, with
\begin{equation}
\vett F^A=\int_V \vett f^A dV
\end{equation}
the Abraham force, with density 
\begin{equation}
\vett f^A=\frac{\partial}{\partial t} \left(\vett D^L\times \vett B-\frac{1}{c^2}\vett E\times \vett H\right)\text{,}
\label{SIabr}
\end{equation}
and $\mathbf{D}^{L}=\epsilon_0 \vett{E}+\vett{P}$ the linear displacement vector.

When averaged versus time the contribution of $\vett F^{A}$ vanishes, and the force is due to the time
average of $\vett F_M$, which after integration by parts, is also written as \cite{Gordon73} 
\begin{equation}
\vett F_M=\int_V \vett{f}_M dV=\int_V \left(-\frac{E^2}{2} \nabla\epsilon \right) dV\text{.}
\label{FL3}
\end{equation}
In terms of the Maxwell stress tensor $\overline{\vett{S}}^L$ calculated on the surface $\Sigma$ of the volume $V$:
\begin{equation}
\vett F_L=\int_\Sigma  \overline{\vett{S}}^L\cdot d\vett A-\frac{d \vett G_A}{d t}
\label{SIFL}
\end{equation}
with 
\begin{equation}
\overline{\vett S}^{L}=\mathbf{D}^{L}\mathbf{E}+\mathbf{B}\mathbf{H}-\frac{\overline{\mathbf I}}{2}\left(\mathbf{D}^{L}\cdot\mathbf{E}+\mathbf{B}
\cdot\mathbf{H}\right) \text{,}
\end{equation}
and the Abraham form of the electromagnetic momentum $\vett{G}_A$, with density $\vett{g}_A$, 
\begin{equation}
\vett G_A=\int_V \vett{g}_A dV=\int \frac{\vett{E}\times\vett{H}}{c^2} dV\text{.}
\end{equation}
The contribution of  $\overline{\vett S}^{L}$ in (\ref{SIFL}) is the temporal derivative of the total momentum:
\begin{equation}
\frac{d \vett{G}_{tot}}{d t}=\int_\Sigma  \overline{\vett{S}}^L\cdot d\vett A\text{.}
\end{equation}
\section{\label{nonlinearforce}The Maxwell stress tensor for nonlinear media}
When including the nonlinearity there is an additional contribution to the force due to the nonlinear polarization; 
this also gives an additional term to the Maxwell stress tensor $\overline{\vett{S}}$, to the Abraham force $\vett{F}^A$ and to mechanical force $\vett{F}_M$. 
The polarization is $\vett{P}=\vett{P}^L+\vett{P}^{NL}$, we also let $\vett{D}=\vett{D}^L+\vett{P}^{NL}$ and $\mathbf{D}^{L}=\epsilon_0 \vett{E}+\vett{P}^L$.
We consider an instantaneous nonlinear response such that
\begin{equation}
\mathbf{P}^{NL}=\epsilon_0\chi^{(3)} \left(\mathbf{E}\cdot \mathbf{E}\right) \mathbf{E}\text{,}
\label{vecnonlinear}
\end{equation}
which, after being written in tensorial notation, reads as (we omit the symbol of summation over repeated indices)
\begin{equation}
P^{NL}_s=\chi_{spqr} E_s E_p E_q E_r
\end{equation}
being
\begin{equation}
\chi_{spqr}=\frac{\epsilon_0\chi^{(3)}}{3}\left(\delta_{sp}\delta_{qr}+\delta_{sq}\delta_{pr}+\delta_{sr}\delta_{pq}\right)
\label{chiindex}
\end{equation}
with $\delta_{ij}$ the Kronecker delta. 
From Maxwell equations
\begin{equation}
\frac{d \mathbf{G}_{tot}}{dt}=\int_{\Sigma} \overline{\vett{S}}^{L} \cdot d\mathbf{A}+\int_V {\bf N} dV\text{.}
\label{generalizedtensor}
\end{equation}
In Eq.(\ref{generalizedtensor}) ${\bf N}$ accounts for $\vett{P}^{NL}$ and is given by 
\begin{equation}
{\bf N}=\left(\nabla\cdot\vett{P}^{NL}\right)\vett{E}+(\nabla\times \vett E)\times \vett{P}^{NL}\text{.}
\end{equation}
Eq.(\ref{generalizedtensor}) holds for a linear medium ($\mathbf{N}=0$), with
\begin{equation}
D_i^{L}=\epsilon_{ij} E_j \text{,}
\end{equation}
and $\epsilon_{ij}=\epsilon_{ji}$  \cite{LandauEDCMbook}.
It is important to show that in the presence of nonlinearity the volume integral in (\ref{generalizedtensor})
can be expressed as an integral over the surface $\Sigma$, and the Maxwell stress tensor can be written
as 
\begin{equation}
\overline{\bf S}=\overline{\bf S}^L+\overline{\bf S}^{NL}
\end{equation}
so that
\begin{equation}
\frac{d\mathbf{G}_{tot}}{dt}=\int_{\Sigma} \overline{\bf S} \cdot d\vett A\text{,}
\label{generalizedtensorfinal}
\end{equation}
holds true also in the presence of nonlinearity.

In this case, the argument of the volume integral in (\ref{generalizedtensor}) has to be a divergence, i.e., ${\bf N}=\nabla\cdot \overline{\vett{S}}^{NL}$. This can be shown by the use of tensorial notation.
Specifically, letting $\nabla_i=\partial/\partial x_i$, and introducing the Levi-Civita symbol $\epsilon_{ijk}$, so that
$\nabla \times \mathbf{E}\cdot \mathbf{\hat{x}}_i=\epsilon_{ijk}\nabla_j E_k$, with $\vett{\hat{x}}_i$ the unit vector in the direction $x_i$, we have 
\begin{equation}
N_i=E_i (\nabla_j P_j^{NL})+\epsilon_{ijk} \epsilon_{jpq} P_k^{NL}\nabla_p E_q\text{.}
\end{equation}
We then use the following well-known identity
\begin{equation}
  \epsilon_{ijk}\epsilon_{ilm}=\delta_{jl}\delta_{km}-\delta_{jm}\delta_{kl}\text{,}
\end{equation}
 and obtain
\begin{equation}
N_i=E_i (\nabla_j P_j^{NL})+P_j^{NL} (\nabla_j E_i)-P_j^{NL} (\nabla_i E_j)\text{.}
\end{equation}
For isotropic materials after (\ref{chiindex}), we have
\begin{equation}
\nabla_i (P_s^{NL} E_s)=\nabla_j (\chi_{spqr} E_s E_p E_q E_r)=4 P_s^{NL} \nabla_i E_s\text{,}
\end{equation}
and finally
\begin{equation}
N_i= (\nabla_j P_j^{NL})E_i-\frac{1}{4} (\nabla_i P_j^{NL} E_j)=\nabla_i S_{ij}^{NL}\text{,}
\end{equation}
with
\begin{equation}
  S_{ij}^{NL}= P_j^{NL}E_i-\frac{1}{4} (P_s^{NL} E_s)\delta_{ij}\text{.}
\end{equation}

In dyadic notation, we have the expression for the nonlinear contribution to the Maxwell stress tensor
\begin{equation}
\overline{\bf S}^{NL}=\mathbf{P}^{NL}\mathbf{E}-\frac{1}{4}(\mathbf{P}^{NL}\cdot \mathbf{E})\overline{\mathbf{I}}=
\epsilon_0 \chi^{(3)} E^2 \mathbf{E}\mathbf{E}-\frac{\epsilon_0 \chi^{(3)} E^4}{4}\overline{\mathbf{I}}\text{,}
\label{SNL}
\end{equation}
being $E^2=\mathbf{E}\cdot\mathbf{E}$.
As observed in \cite{LandauEDCMbook}, for a finite block, the fact that the force can be calculated as
a surface integral is a consequence of momentum conservation, and the use of a surface in vacuum is justified by the continuity of
the forces. However, for an-isotropic, linearly and nonlinearly absorbing, non-homogeneous or more complicated media, 
this may not be satisfied. 

We remark that Eq.(\ref{SNL}) is {\it different} from the expression obtained letting $\mathbf{D}=\epsilon_0 \mathbf{E}+\mathbf{P}^L+\mathbf{P}^{NL}$,
in the standard {\it linear} stress tensor.

\subsection{The nonlinear Abraham force}
The force acting on the medium $\vett{F}$ is the sum of the mechanical force $\vett{F}_M$ and of the Abraham force $\vett{F}^A$, it is given by the time derivative of the total momentum minus the momentum of the EM field:
\begin{equation}
\vett{F}=\vett{F}^A+\vett{F}_M=\frac{d \vett{G}_{tot}}{dt}-\frac{d \vett{G}_{A}}{dt}\text{.}
\end{equation}

For a linear medium, the Abraham force density has the known expression (\ref{SIabr}), which is rewritten as
\begin{equation}
\mathbf{f}^{AL}=\frac{\partial}{\partial t}\left(\mathbf{D}^L\times\mathbf{B}-\frac{1}{c^2}\mathbf{E}\times\mathbf{H}\right)\text{.}
\end{equation}
In the presence of a nonlinear polarization the total Abraham force is 
\begin{equation}
\mathbf{f}^{A}=\mathbf{f}^{AL}+\mathbf{f}^{ANL}\text{,}
\end{equation}
with a nonlinear contribution given by
\begin{equation}
\mathbf{f}^{ANL}=\frac{\partial}{\partial t}\left(\mathbf{P}^{NL}\times\mathbf{B}\right)\text{.}
\end{equation}
In the specific case of an isotropic instantaneous nonlinearity, and for a linearly polarized plane wave propagating
in the $z-$direction (with unit vector $\vett{\hat{z}}$), being $I$ the optical intensity, we have 
\begin{equation}
\mathbf{f}^{ANL}=\hat{\mathbf{z}}\frac{\chi^{(3)}}{c}\frac{\partial I^4}{\partial t}\text{.}
\end{equation}
As for the linear case, when averaged w.r.t. time the Abraham force vanishes, and 
does not contribute to the pressure on the block.
\subsection{The nonlinear mechanical force density}
The total force density is written as
\begin{equation}
\vett{f}=\vett{f}_L+\vett{f}_{NL}\text{,}
\end{equation}
where $\vett{f}_L=\vett{f}^{AL}+\vett{f}_{ML}$ is given above and
\begin{equation}
\vett{f}_{NL}=\vett{f}^{ANL}+\vett{f}_{MNL}\text{.}
\end{equation}
In isotropic media $\vett{f}_{MNL}$ can be written as
\begin{equation}
\vett{f}_{MNL}=\frac{\chi^{(3)}}{4}\nabla(E^4)\text{.}
\end{equation}
The total force density is 
\begin{equation}
\vett{f}=\vett{f}^A+\vett{f}_M
\end{equation}
and 
\begin{equation}
\vett{f}_M=\frac{\epsilon_0 (n^2-1)}{2}\nabla E^2+\frac{\epsilon_0 \chi^{(3)}}{4}\nabla E^4\text{,}
\end{equation}
or equivalently, neglecting surface effects and by integration by parts in the relevant volume integral:
\begin{equation}
\vett{f}_M=-\frac{E^2}{2}\nabla\epsilon_L-\frac{\epsilon_0 E^4}{4}\nabla{\chi^{(3)}}{,}
\end{equation}
with $\epsilon_L=\epsilon_0 (1+\chi^{1})$ the linear susceptibility.
\section{\label{delayed} Delayed nonlinear response}
We consider a non-instantaneous nonlinear response, which we introduce in our model  
by writing in Eq.(\ref{iLcw}): $\phi_{NL}=2\pi n_2 I(0,t-t_{NL})L/\lambda$, with $t_{NL}$ the delay-time of the nonlinear phase-shift.
By repeating the analysis in the main text we have that Eq.(\ref{p2cw}) becomes
\begin{equation}
p^{CW}_2= \frac{\pi \mathcal{T}  n_2 L }{c \lambda}  
\sin\left[2\omega \left(\frac{n_0 L}{c} - t_{NL}\right)\right]\text{,}
\label{p2cwTnl}
\end{equation}
which shows that a delay in the nonlinear optical response may cause a spectral shift of the
nonlinear pressure coefficient with respect to the instantaneous case.
\section{\label{pulsedall} $p_2$ in the pulsed regime}
Here we report the full expression for $p_2$ as obtained in the case of pulsed excitation:
\begin{widetext}
\begin{equation}
p_2=\frac{\pi L n_2 \mathcal{T}e^{-\frac{L^2 n_g^2}{2 c^2 T_0^2}}}{\lambda c}\left\{
\sin\left(\frac{4\pi L n_0}{\lambda}\right)+
e^{-\frac{4\pi^2 T_0^2}{T_{opt}^2}}\sin\left[\frac{4\pi L (n_0-n_g)}{\lambda}\right]+
e^{-\frac{2\pi^2 T_0^2}{T_{opt}^2}}\sin\left[\frac{2\pi L (2n_0-n_g)}{\lambda}\right]
\right\}\text{,}
\label{hugep2}
\end{equation}
\end{widetext}
with $n_2$,$n_0$, $\mathcal{T}$, and $n_g$ dependent of $\lambda$, 
$T_{opt}=c/\lambda$ is the optical cycle, and $n_g=v_g/c$ is the group index. 
Note that Eq.(\ref{hugep2}) includes terms that are very small when the pulse duration contains
few optical cycles, when $T_{0}>>T_{opt}$, in this limit Eq.(\ref{p2pulse})
in the main text is derived from Eq.(\ref{hugep2}).
\section{\label{codedetail} Details on the numerical code and of the adopted dispersion relation in the linear and nonlinear case}
In the numerical simulations we include both the material dispersion for the linear response, and for
the nonlinear susceptibility. This is done following the standard textbook approach for describing the
nonlinear response of electronic nonlinearity in which the Maxwell equations are coupled to a nonlinear oscillator equation\cite{TafloveBook}. The Maxwell equations are written as
\begin{equation}
\begin{array}{l}
\nabla\times\mathbf{E}=-\mu_0 \partial_t\mathbf{H}\\
\nabla\times\mathbf{H}=\epsilon_0 \partial_t \mathbf{E}+\partial_t \mathbf{P}\text{,}
\end{array}
\end{equation}
with $\mathbf{P}$ the material polarization including the linear and the nonlinear part.
$\mathbf{P}$ obeys to the second order equation
\begin{equation}
\frac{d^2 \mathbf{P}}{dt^2}+2\gamma \frac{d \mathbf{P}}{dt}+f(P)\omega_0^2 \mathbf{P}=\omega_0^2 (\epsilon_r-1)\epsilon_0 \mathbf{P}\text{.}
\label{polarizationequation}
\end{equation}
Note that Eq.(\ref{polarizationequation}) is used in the spatial locations where the block is present, otherwise $\mathbf{P}=0$, corresponding to vacuum. 
In regions where the material is present $\epsilon_r$, $\omega_0$, and $\gamma$ are coefficients determining the
linear dispersion. $f(P)$ is a function of the modulus $P=(\mathbf{P}\cdot\mathbf{P})^{1/2}$.
For a linear medium $f(P)=1$ and Eq.(\ref{polarizationequation}) corresponds to a single pole oscillator, which mo\-dels a linear dispersive
medium with dispersion relation
\begin{equation}
\chi^{(1)}(\omega)=\frac{\omega_0^2 (\epsilon_r-1)}{-\omega^2-2 i \omega \gamma +\omega_0^2}\text{.}
\label{lineardispersion}
\end{equation}
We choose $\epsilon_r=2.1045$, $\omega_0=7\times 10^{15}$~rad/s, furnishing the linear dispersive refractive index $n_0(\lambda)$ in figure \ref{figuretheory}A. Note that we also include losses in the model with $\gamma=7\times 10^{13}$~s$^{-1}$,
resulting in a linear transmission, due to absorption, of about $90\%$ from the block at $\lambda=800$~nm;
i.e., we include in the simulations a not negligible amount of linear losses, which is not present in the theoretical data in Figure \ref{figuretheory}A.

The isotropic nonlinear response is obtained, in the simplest formulation, by writing $f(P)=1+\chi^{(3)} P^3$, with $\chi^{(3)}$ a material dependent coefficient that determines the frequency dependent 
Kerr coefficient $n_2$. 
The cubic $f(P)$ is indeed an approximation for more general models;
in our code we use the function $f(P)=1/(1+\chi^{(3)} P^3)^{3/2}$, which, at the lowest order in $\chi^{(3)}$ is equivalent to the cubic function, but also includes higher order nonlinearity.\cite{Koga99}

By using standard perturbation theory (as reported in many textbooks, as, e.g.,\cite{BloembergenBook}) it is possible to write for this model
\begin{equation}
n_2(\omega)=\frac{9\epsilon_0 \chi^{(3)} \chi^{(1)}(\omega)^4 }{8c (\epsilon_r-1)n_0(\omega)}
\end{equation}
so that $n_2$ is directly proportional to the strength of the nonlinear coefficient $\chi^{(3)}$, and it is also frequency dependent as shown in Fig.\ref{figuretheory}A.

We stress that this approach is more realistic than FDTD codes based on
iterative algorithms with instantaneous nonlinearity (for a discussion see, e.g., \cite{TafloveBook}), and it also accounts for the dispersion of the nonlinear coefficients and satisfies the relevant Kramers-Kronig relations for the causality of linear and nonlinear response.
Our parallel code is a C++ 3D+1 FDTD based on the MPI-II protocol and running on the FERMI IBM Blue Gene Q system at CINECA, within the Italian Supercomputing Resource Allocation (ISCRA) initiative.

%

\end{document}